\documentclass[12pt,letterpaper]{article}

\usepackage[margin=1in]{geometry}
\usepackage{graphicx}
\usepackage{amsmath}
\usepackage{amssymb}
\usepackage{multirow}
\usepackage{graphicx}
\usepackage{color}
\usepackage{tabularx}
\usepackage{setspace}
 \usepackage{hyperref}
 \usepackage{algorithm}
\usepackage{algpseudocode}
\usepackage{caption}
\usepackage{lscape}
\usepackage{rotating}
\usepackage{calligra}
\numberwithin{equation}{section}
\usepackage[nomarkers,nolists]{endfloat}

\title{A pseudo-outcome-based framework to analyze treatment heterogeneity in survival data using electronic health records}
\author{
    Na Bo\textsuperscript{1}, Jong-Hyeon Jeong\textsuperscript{1,2}, Erick Forno\textsuperscript{1,3}, Ying Ding\textsuperscript{1,*}
}
\date{}

\begin{document}
\maketitle

\begin{center}
    \textsuperscript{1}Department of Biostatistics, University of Pittsburgh, Pittsburgh, PA, U.S.A.\\
    \textsuperscript{2}Division of Cancer Treatment and Diagnosis, National Cancer Institute, MD, U.S.A\\
    \textsuperscript{3}Department of Pediatrics, Indiana University School of Medicine, IN, U.S.A.\\
    \textsuperscript{*}email: yingding@pitt.edu
\end{center}

\abstract{An important aspect of precision medicine focuses on characterizing diverse responses to treatment due to unique patient characteristics, also known as heterogeneous treatment effects (HTE), and identifying beneficial subgroups with enhanced treatment effects. Estimating HTE with right-censored data in observational studies remains challenging. In this paper, we propose a pseudo-outcome-based framework for analyzing HTE in survival data, which includes a list of meta-learners for estimating HTE, a variable importance metric for identifying predictive variables to HTE, and a data-adaptive procedure to select subgroups with enhanced treatment effects. We evaluate the finite sample performance of the framework under various settings of observational studies. Furthermore, we applied the proposed methods to analyze the treatment heterogeneity of a Written Asthma Action Plan (WAAP) on time-to-ED (Emergency Department) return due to asthma exacerbation using a large asthma electronic health records dataset with visit records expanded from pre- to post-COVID-19 pandemic. We identified vulnerable subgroups of patients with poorer asthma outcomes but enhanced benefits from WAAP and characterized patient profiles. Our research provides valuable insights for healthcare providers on the strategic distribution of WAAP, particularly during disruptive public health crises, ultimately improving the management and control of pediatric asthma.}


\section{Introduction}
\label{sec:intro}
Traditionally, treatment effectiveness is investigated through randomized clinical trials (RCTs) to estimate the average treatment effects (ATE), which often overlook heterogeneous treatment effects (HTE) based on unique patient characteristics. Nowadays, routinely collected healthcare databases, such as electronic health records (EHR), provide a comprehensive dataset including patient characteristics and medical records with large sample sizes. By leveraging this wealth of information, researchers can better identify subgroups of patients who may exhibit differential responses to specific treatments, thus enhancing the precision of medical interventions. 

HTE is typically defined through the conditional average treatment effect (CATE), which refers to the difference in the conditional expectation of the outcome given covariates between treatment and control groups. Recently, there has been growing attention to estimating the HTE in survival outcomes, which is particularly crucial in chronic disease care as healthcare providers continually strive to seek individualized patient care based on long-term clinical outcomes such as disease progression trajectories. 
The existing methods for estimating CATE in survival data contain two broad categories: directly estimating CATE and indirectly estimating CATE. 
Meta-learners are a group of two-step algorithms to directly estimate CATE, with the first step estimating the nuisance parameters and the second step estimating or calculating CATE by plugging in those estimated nuisance parameters. For example, Xu et al. \cite{metalearner_survival_casestudy_Xu2023,CATEsurvTutorialBookChap_Xu2022_inbook} and Bo et al. \cite{metalearnerSurv_Bo2023} estimated CATE under the survival probability scale using various meta-learners in an RCT setting. Tree-based methods are another popular approach to estimating CATE directly. For example, the Bayesian Accelerated Failure Time model (BAFT) estimates CATE through Bayesian additive regression trees, which can quantify the uncertainty of estimated CATE by posterior intervals \cite{BAFT_Henderson2018_biostatistics, HTE_mediansurv_ML_Hu2021, clusteredCATEsurv_Hu2022}. More recently, Cui et al. \cite{CSF_JRSSb2023} proposed causal survival forests (CSF) by solving orthogonal estimating equations. 
For indirectly estimating CATE, targeted minimum loss-based estimation has been extended to estimate (pre-specified) subgroup-specific treatment effect under the survival probability scale \cite{TMLE_survival_Zhu_JBI2020}.
Representation learning, employed for learning potential outcomes, indirectly estimates CATE by calculating the difference between predicted potential outcomes. 
For example, Curth et al. \cite{survITE_NeurIPS2021_Curth} introduced survITE to estimate individualized hazard functions at discretized survival times under each treatment. 

Our study is motivated by an investigation into the effectiveness of pediatric asthma care using a large EHR database. This complex and multifactorial chronic disease poses a major public health challenge, affecting 4.7 million children in the U.S. and resulting in 1.8 million asthma attacks and 270,330 emergency department (ED) visits annually. 
Poorly controlled asthma is a significant source of morbidity, especially in young children and those with severe diseases. The National Heart, Lung, and Blood Institute recommends Written Asthma Action Plans (WAAP) as a pivotal intervention for asthma control. WAAPs provide personalized guidelines based on individual medical history and symptoms, aiming to reduce unscheduled healthcare visits for asthma exacerbation.  
Recent studies that evaluated the effectiveness of WAAP using EHR data primarily focus on WAAP's effect in reducing the number of unscheduled visits \cite{cross_section_survey_AAP_Alkhthlan2021, Var_ChildrenAsthmaED_EHR_Shechter2019}. However, research into WAAP's effect on survival outcomes, such as time-to-ED return due to asthma exacerbation (abbreviated as ``time-to-ED return"), lacks studies, with only a few conducted before the COVID-19 pandemic through RCTs and showing controversial results. For example, one RCT found a significantly lower cumulative incident rate in time-to-ED return in the WAAP group than the control group \cite{ED_WAAP_RCT_HighMorbidity_children_Teach2006}, while another similar RCT found no significant difference \cite{AsthmaEducation_RCT_BROWN2006}. These conflicting findings from small sample size RCTs underscore the necessity to investigate the HTE of WAAP further using large observational data, such as EHR. Additionally, the influence of the COVID-19 pandemic on pediatric asthma care remains unclear, warranting additional research to investigate WAAP’s effectiveness during disruptive public health events.

In this paper, we propose a pseudo-outcome-based framework for analyzing HTE in survival data, which includes six meta-learners to estimate CATE under the survival probability scale, an evaluation process to quantify the predictability of covariates and a data-adaptive strategy to select beneficial subgroups with enhanced treatment effects.
The meta-learners offer several attractive advantages. 
For example, they do not need to discretize the survival time, have flexibility in both nuisance parameters and CATE function estimations, allow the application of various machine learning or model-based regression methods, and directly target the CATE function to make the interpretation easier. We utilized SHAP values \cite{SHAP_2017} to quantify variable importance (VIP), which gives individualized VIP scores for each variable across each subject. Furthermore, the data-adaptive strategy can dynamically examine the treatment heterogeneity across the entire study population, which helps identify beneficial subgroups. 

The remainder of this paper is organized as follows. In section \ref{sec:data_explore}, we describe the pediatric asthma EHR data with an exploratory analysis. Section \ref{sec:method} presents our proposed framework for analyzing HTE in survival data. In section \ref{sec:simu}, comprehensive simulations are conducted under both RCT and observational study settings. In section \ref{sec:realdat_application}, we implement the procedure to analyze the heterogeneous effects of WAAP on time-to-ED return among pediatric asthma patients under the COVID-19 disruption. Section \ref{sec:discuss} concludes our proposed procedure for analyzing HTE and discusses the limitations and future directions.

\section{Pediatric Asthma EHR Data}
\label{sec:data_explore}
In this study, we analyzed EHR data of pediatric asthma patients (2-21 years old) obtained from the Children's Hospital of Pittsburgh (IRB: STUDY22040043). The dataset curated 32,000+ encounters (i.e., visits) from 13,000+ patients between January 2019 and January 2023. Among them, over 8,700 encounters are ED visits due to asthma exacerbation. Our aim was to assess the impact of WAAP on the time-to-ED return (due to asthma exacerbation) under the COVID-19 disruption. The time origin for each individual was set to be their first ED visit in our curated EHR data (denoted as the ``index ED visit''). The time-to-ED return was defined as the time (in days) from their index ED visit to the next ED visit or last visit (if no additional ED visit occurred, i.e., treated as censored). Those who had only one ED visit but no additional visits were excluded from this study. The analysis dataset contains 1,903 patients, with 1,368 patients (70\%) having an ED return. A total of 471 patients (25\%) received WAAP. The baseline covariates include sex (female and male), race (black, white, others), existence of chronic diseases (bronchomalacia, bronchopulmonary dysplasia, immunodeficiency, pulmonary hypertension or tracheomalacia), existence of acute respiratory diseases (acute bronchospasms or subcutaneous emphysema), admission to hospital at their index ED visit (Yes or No), influenza vaccination (Yes or No), and age. In addition, the index ED visit was categorized into the pre-pandemic period (January 1, 2019 to March 14, 2020), pandemic period (March 15, 2020, to April 30, 2021), and post-pandemic period (after May 1, 2021). The cut-off dates for the three periods are based on the COVID-19 lockdown date and the major back-to-school date.

Table \ref{tab:asthma_baseline_cov_character} summarizes patient baseline characteristics by WAAP (Yes or No). Age, acute respiratory disease, and hospitalization rates significantly differ between those with and without WAAP ($p<0.05$). For example, younger age groups have more access to WAAP compared to the adolescent age group (25\%-26\% of 2-12 years were given WAAP, while only 17.5\% of 13-21 years were given WAAP).
The distribution of WAAP or not also differs significantly across pandemic periods ($p=0.017$), with less WAAP given during- and post-pandemic compared to pre-pandemic (23.3\% and 21.7\% vs 27.6\%). We also observe the race disparities in accessing the WAAP, with white patients being given WAAP more frequently than black patients (27.0\% white were given WAAP while 22.7\% black were given WAAP).

\begin{table}
\caption {Patient baseline characteristics of pediatric asthma EHR data, overall and by WAAP.}
\label{tab:asthma_baseline_cov_character}
\centering
\scalebox{0.8}{
\begin{tabular}{lllll}
\hline
& \begin{tabular}{@{}l@{}}All \\ ($n=1,903$)\end{tabular}     & \begin{tabular}{@{}l@{}}No WAAP \\ ($n=1,432$)\end{tabular}& \begin{tabular}{@{}l@{}}WAAP \\ ($n=471$)\end{tabular} & \textit{p}-value$^\star$ \\
\hline
Age (n, \%$^{\star \star}$)                                                                           &                  &                  &                     & 0.033  \\
\multicolumn{1}{c}{2-4 years old} & 899 (47.2)       & 672 (74.7)       & 277 (25.3)          &         \\
\multicolumn{1}{c}{5-12 years old} & 793 (41.7)       & 586 (73.9)       & 207 (26.1)          &         \\
\multicolumn{1}{c}{13-21 years old} & 211 (11.1)       & 174 (82.5)       & 37 (17.5)          &         \\
Sex (n, \%)                                                                   &                  &                  &                     &  0.263     \\
\multicolumn{1}{c}{Female} & 767 (40.3)       & 588 (76.7)       & 179 (23.3)          &         \\
\multicolumn{1}{c}{Male} & 1136 (59.7)       & 844 (74.3)       & 292 (25.7)          &         \\
Race (n, \%)                                                               &                  &                  &                     & 0.093     \\
\multicolumn{1}{c}{Black} & 851 (44.7)       & 658 (77.3)       & 193 (22.7)          &         \\
\multicolumn{1}{c}{White} & 907 (47.7)       & 662 (73.0)       & 245 (27.0)          &         \\
\multicolumn{1}{c}{Others} & 145 (7.6)       & 112 (77.2)       & 33 (22.8)          &         \\
Chronic disease existence (n, \%)      &                  &                  &                  &    0.113   \\
\multicolumn{1}{c}{True} & 79 (4.2)       & 53 (67.1)       & 26 (32.9)          &         \\
\multicolumn{1}{c}{False} & 1824 (95.8)       & 1379 (75.6)       & 445 (24.4)          &         \\
Acute respiratory disease existence (n, \%)      &                  &                  &                  &    0.008   \\
\multicolumn{1}{c}{True} & 24 (1.3)       & 12 (50.0)       & 12 (50.0)          &         \\
\multicolumn{1}{c}{False} & 1879 (98.7)       & 1420 (75.6)       & 459 (24.4)          &         \\
Influenza vaccination (n, \%)      &                  &                  &                  &    0.148   \\
\multicolumn{1}{c}{True} & 1030 (54.1)       & 761 (73.9)       & 269 (26.1)          &         \\
\multicolumn{1}{c}{False} & 873 (45.9)       & 671 (76.9)       & 202 (23.1)          &         \\
Hospitalization (n, \%)              &                  &                  &                  & $<.001$\\
\multicolumn{1}{c}{True} & 915 (48.1)       & 455 (49.7)       & 460 (50.3)          &         \\
\multicolumn{1}{c}{False} & 988 (51.9)       & 977 (98.9)       & 11 (1.1)          &         \\
Pandemic period (n, \%)      &                  &                  &                  &    0.017   \\
\multicolumn{1}{c}{Pre-pandemic} & 923 (48.5)       & 668 (72.4)       & 255 (27.6)          &         \\
\multicolumn{1}{c}{During-pandemic} & 219 (11.5)       & 168 (76.7)       & 51 (23.3)          &         \\
\multicolumn{1}{c}{Post-pandemic} & 761 (40.0)       & 596 (78.3)       & 165 (21.7)          &         \\
\hline
\multicolumn{5}{l}{\small $^\star p$-value was computed using Chi-squared tests.}\\ 
\multicolumn{5}{l}{\small $^{\star \star}$The percentage of each covariate by WAAP groups was calculated as the row percent.}\\
\end{tabular}}
\end{table}

\begin{figure}[ht]
\centering
\includegraphics[width=5.5in]{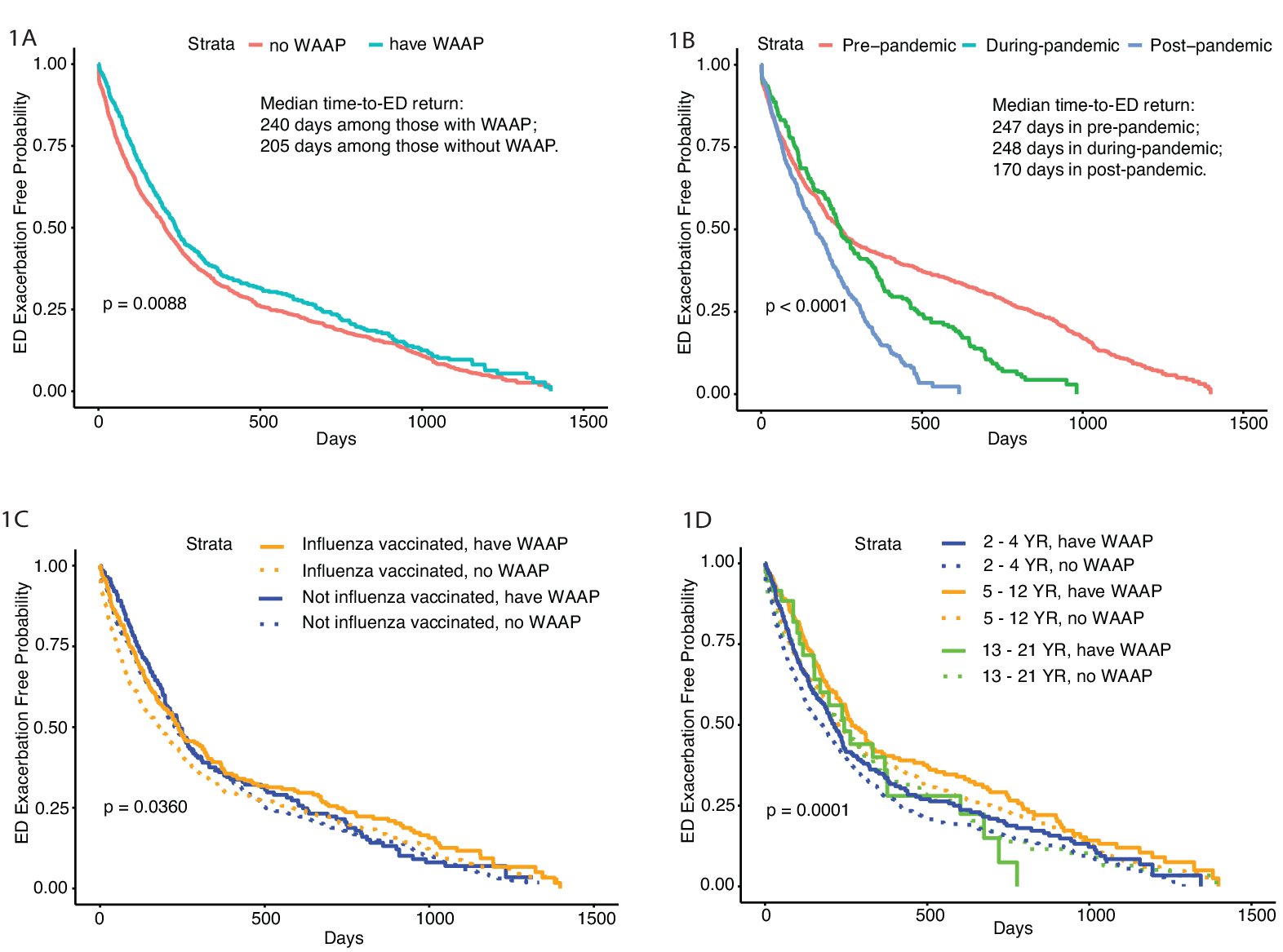}
\caption{Figure 1A and 1B present Kaplan Meier (KM) curves by WAAP intervention groups and by pandemic periods. Figures 1C and 1D present KM curves of WAAP interaction with the influenza vaccine and age. Log-rank tests were conducted to compare the survival curves across different categories.}
\label{fig:asthma_KM_WAAP_pandemic3cat}
\end{figure}

Figure \ref{fig:asthma_KM_WAAP_pandemic3cat}.A shows that patients without WAAP returned to ED sooner compared to patients with WAAP ($p=0.0088$), but the separation of Kaplan Meier (KM) curves is marginal. Figure \ref{fig:asthma_KM_WAAP_pandemic3cat}.B displays significant differences in KM curves across three pandemic periods. Patients experienced faster ED return in post-pandemic than pre- and during-pandemic periods. Moreover, we observe potential subgroup effects of WAAP. For example, in Figure  \ref{fig:asthma_KM_WAAP_pandemic3cat}.C, among patients vaccinated with the influenza vaccine, those who have WAAP show a longer time-to-ED return compared to those without WAAP ($p=0.013$). Conversely, we do not observe significant differences in time-to-ED return between the two treatment groups among those without influenza vaccination. Additionally, as shown in Figure \ref{fig:asthma_KM_WAAP_pandemic3cat}.D, among younger patients (2-4 years old and 5-12 years old), those with WAAP exhibit a longer time-to-ED return than those without WAAP ($p=0.0068$ in 2-4 years; $p=0.072$ in 5-12 years). However, there is no obvious difference in time-to-ED return between the two treatment groups in 13-21 years. 

Our exploratory analysis reveals subgroup heterogeneity in the effectiveness of WAAP on time-to-ED return among pediatric asthma patients, underscoring the need for additional investigation into the HTE of WAAP. Furthermore, the pandemic period demonstrates strong associations with both the outcome and the treatment, which prompts us to include the pandemic period in assessing the effectiveness of WAAP on time-to-ED return.

\section{Proposed Method for Analyzing HTE in Survival Data}
\label{sec:method}

\subsection{Notation, definition, and assumptions}
\label{sec:method_notation}
Consider an observational study or RCT of sample size $n$, which compares two treatment arms, $A \in \{0, 1\}$. $\boldsymbol{X}$ is a $p-$dimensional vector of baseline covariates; $\boldsymbol{x}$ denotes a realization of the covariate vector $\boldsymbol{X}=\boldsymbol{x}$; $\boldsymbol{X}_i$ denotes the covariate vector for subject $i$, $i=1, 2, \dots, n$; $X_{ij}$ denotes the $j$th covariate for the individual with covariate $\boldsymbol{X}_i$. Let $T$ denote the survival time and $C$ denote the censoring time. Then we define $U=\min(T,C)$ as the observed time with event indicator $\delta$ defined as $\delta=I(T<C)$.  The observed data is denoted as $D = \{(U_i, \delta_i), A_i, \boldsymbol{X}_i; i=1, \dots, n \}$. 
In this paper, we follow the Neyman-Rubin potential outcome framework \cite{rubin1974,Neyman1990} to define CATE in survival data. Let $T(1)$ and $T(0)$ denote potential survival times for a patient receiving treatment $A=1$ and $A=0$, respectively. Similarly, $C(1)$ and $C(0)$ denote potential censoring times and $\delta(1)$ and $\delta(0)$ denote potential event indicators. 
We define CATE as the difference of survival probabilities at a given time $t^*$ between two treatment arms, given baseline covariates:
\begin{eqnarray}
\label{eq:cate_def_survp}
\tau(\boldsymbol{x};t^*)&=&E[I(T(1)>t^*)-I(T(0)>t^*)|\boldsymbol{X}=\boldsymbol{x}]  \nonumber \\ 
&=&E[I(T>t^*)|A=1, \boldsymbol{X}=\boldsymbol{x}]-E[I(T>t^*)|A=0,\boldsymbol{X}=\boldsymbol{x}]  \nonumber \\
&=&S_1(t^*|\boldsymbol{x})-S_0(t^*|\boldsymbol{x}) .
\end{eqnarray}

The first line of definition (\ref{eq:cate_def_survp}) is the definition of CATE under Neyman–Rubin potential outcome framework. Note that the second line of definition (\ref{eq:cate_def_survp}) can be proved by assuming the three common causal inference assumptions: consistency, unconfoundedness and positivity. Consistency assumption means that the potential outcome under treatment $A=a$ is equal to the observed outcome if the actual treatment $a$ is recevied, denoted as $T=AT(1)+(1-A)T(0)$, $C=AC(1)+(1-A)C(0)$ and $\delta=A\delta(1)+(1-A)\delta(0)$. The unconfoundedness assumes no unmeasured confounders, written as $(T(0), T(1)) \!\perp\!\!\!\perp A|\boldsymbol{X}$. The positivity assumption requires that there is a positive probability of being assigned to either treatment for each value of covariates: $P(A=1|\boldsymbol{X}=\boldsymbol{x}) \in (0,1)$. We also assume the non-informative censoring assumption under which the censoring time is independent of the survival time given covariates and treatment assignment: $C(a)\!\perp\!\!\!\perp T(a) | (\boldsymbol{X}, A)$.

\subsection{A pseudo-outcome-based framework to estimate CATE}
\label{sec:method_pseudoreg}
In this paper, we directly estimate CATE by constructing an objective function, where the survival functions are treated as nuisance parameters. In addition, we model treatment assignment probability conditioning on covariates. If we have access to individualized treatment effects (ITE), denoted as $Y_i=I(T_i(1)>t^*)-I(T_i(0)>t^*)$, $\tau(\boldsymbol{x};t^*)$ can be directly estimated by regressing ITEs (i.e., $Y_i$) on covariates. However, such ITEs cannot be obtained due to two reasons: $I(T>t^*)$ cannot be known for those censored before time $t^*$, and even in the absence of censoring, patients cannot receive both treatment and control simultaneously, making one of $T(1)$ and $T(0)$ unobserved. Instead, we construct a pseudo individualized treatment effect (pITE) $Y^*$ (e.g., a pseudo-outcome) and estimate $\tau(\boldsymbol{x};t^*)$ by minimizing a squared error loss: 
\begin{eqnarray}
\label{loss:squared_error_no_censor}
    \frac{1}{n^o} \sum_{i=1}^{n^o} w_i^M\left(Y_i^*-\tau(\boldsymbol{X}_i;t^*)\right)^2,
\end{eqnarray}
where $n^o$ denotes the sample size of complete data ${D}^o$ whose $I(T>t^*)$ is known (i.e., individuals who are censored before $t^*$ are excluded). $Y_i^*$ and $w_i^M$ are method-specific, which will be introduced in the following paragraphs. To accommodate right-censored data, we propose to incorporate inverse probability censoring weighting (IPCW) to the squared error loss (\ref{loss:squared_error_no_censor}), denoted as $w_i^C=\frac{1}{P(C> \min(T_i,t^*)|\boldsymbol{X}_i,A_i)}$, to predict the inverse of the probability for not being censored before $\min (T_i, t^*)$ for those with $I(T>t^*)$ being known. Therefore, our final squared error loss becomes 
\begin{eqnarray}
    \frac{1}{n^o} \sum_{i=1}^{n^o} w_i^C w_i^M\left(Y_i^*-\tau(\boldsymbol{X}_i;t^*)\right)^2.
\end{eqnarray} 

A group of meta-algorithms can be considered to construct pITE $Y^*$ with a learner-specific weight $w^M$, namely, ``pseudo-outcome-based meta-learners". We introduce six such meta-learners in the following subsection.

\subsubsection{Pseudo-outcome-based meta-learners}
\label{sec:method_metalearner_explain}
\hfill \break
\indent \textbf{X-learner}. The X-learner was originally proposed by Kunzel et al. \cite{pnas_metalearner2019} and consists of two steps. In step 1, the pITE $Y^*_{XL}$ (the subscript $XL$ here represents the X-learner) is constructed by taking the difference between the observed outcome and predicted counterfactual survival probability: $Y^*_{XL}=I(T>t^*)-\hat{S}_0(t^*|\boldsymbol{X})$ for $A=1$ and $Y^*_{XL}=\hat{S}_1(t^*|\boldsymbol{X})-I(T>t^*)$ for $A=0$.
$\hat{S}_0(t^*|\boldsymbol{X})$ and $\hat{S}_1(t^*|\boldsymbol{X})$ are predicted counterfactual survival probabilities on the ``unassigned'' treatment arm. 
In step 2, the treatment-specific CATE functions $\tau_0(\boldsymbol{x};t^*)$ and $\tau_1(\boldsymbol{x};t^*)$ are estimated by regressing $Y^*_{XL}$ on covariates using the subjects in each treatment arm separately. The final CATE is a weighted average of the treatment-specific CATE $\tau(\boldsymbol{x};t^*)=e(\boldsymbol{x})\tau_0(\boldsymbol{x};t^*)+(1-e(\boldsymbol{x}))\tau_1(\boldsymbol{x};t^*)$, where $e(\boldsymbol{x})=P(A=1|\boldsymbol{X}=\boldsymbol{x})$ is the propensity score of receiving treatment, which is usually unknown and needs to be estimated for observational studies. Note that the learner-specific weight $w^M$ is one for all subjects.

\textbf{M-learner and DR-learner}.The M-learner was inspired by the inverse probability weighting (IPW) estimator \cite{IPW1952_Horvitz_Thompson}. 
Specifically, the pITE is constructed as 
\begin{eqnarray}
    Y^*_{ML}=\frac{AI(T>t^*)}{\hat{e}(\boldsymbol{X})}-\frac{(1-A)I(T>t^*)}{1-\hat{e}(\boldsymbol{X})} = \frac{A-\hat{e}(\boldsymbol{X})}{\hat{e}(\boldsymbol{X})\left(1-\hat{e}(\boldsymbol{X})\right)}I(T>t^*). \nonumber
\end{eqnarray}
While the doubly-robust learner (DR-learner), proposed by Kennedy \cite{kennedyDR2022}, was based on the efficient influence curve to estimate ATE, in which 
\begin{eqnarray}
    Y^*_{DR}=\frac{A-\hat{e}(\boldsymbol{X})}{\hat{e}(\boldsymbol{X})\left(1-\hat{e}(\boldsymbol{X})\right)}\left(I(T>t^*)-\hat{S}_A(t^*|\boldsymbol{X})\right)+\hat{S}_1(t^*|\boldsymbol{X})-\hat{S}_0(t^*|\boldsymbol{X}). \nonumber
\end{eqnarray} 
DR-learner has the doubly-robust feature, i.e., the CATE estimator is consistent as long as either the propensity score or the outcome regression models are estimated consistently. Both M- and DR-learner satisfy $\tau(\boldsymbol{x};t^*)=E[Y^*|\boldsymbol{X}=\boldsymbol{x},t^*]$. The learner-specific weight $w^M$ is also one for all subjects. CATE is then estimated by regressing pITE on the covariates.

\textbf{D-learner and DEA-learner}. Lu et al. \cite{Dlearning_TianLu2014} proposed a direct learning method, namely, the D-learner, which does not need to estimate the conditional mean outcome model. The idea is to re-weight covariates which leads to minimize an objective function: 
\begin{eqnarray}
    \frac{1}{n}\sum_{i=1}^n (2A_i-1) \frac{A_i-e\left(\boldsymbol{X}_i\right)}{4e\left(\boldsymbol{X}_i\right)\left(1-e\left(\boldsymbol{X}_i\right)\right)}\left\{2(2A_i-1) I(T_i>t^*)-\tau\left(\boldsymbol{X}_i; t^*\right)\right\}^2. \nonumber
\end{eqnarray}
Thus, we can get D-learner's pITE as $Y_{DL}^*=2(2 A-1) I\left(T>t^*\right)$ and the learner-specific weight $w^M=(2A-1) \frac{A-e\left(\boldsymbol{X}\right)}{4e\left(\boldsymbol{X}\right)\left(1-e\left(\boldsymbol{X}\right)\right)}$.

Chen et al. \cite{DEAlearning2017} extended D-learner to DEA-learner by replacing the observed outcome
$I(T>t^*)$ with its residual $I(T>t^*)-\hat{S}(t^*|\boldsymbol{X})$. The pITE for DEA-learner is $Y_{D E A}^*=2\left(2 A-1\right)\left(I\left(T>t^*\right)-\hat{S}(t^*|\boldsymbol{X})\right)$, where $\hat{S}(t^*|\boldsymbol{X})$ is the predicted survival probability using all subjects (ignore the treatment). The weight is the same as D-learner.

\textbf{R-learner}. Nie and Wager \cite{Rlearner2021} proposed to estimates CATE through the Robinson decomposition: $ I(T_i>t^*)-E\left[I(T>t^*)|\boldsymbol{X}=\boldsymbol{x}, t^*\right]=\left(A_i-e\left(\boldsymbol{X}_i\right)\right) \tau\left(\boldsymbol{X}_i;t^*\right)+\varepsilon_i$,
where $E\left[\varepsilon_i|\boldsymbol{X}_i, A_i\right]=0$. Thus, CATE can be estimated by minimizing the objective function 
\begin{eqnarray}
    \frac{1}{n} \sum_{i=1}^{n}\left(A_i-e\left(\boldsymbol{X}_i\right)\right)^2\left\{\frac{I(T_i>t^*)-S\left(t^*|\boldsymbol{X}_i\right)}{A_i-e\left(\boldsymbol{X}_i\right)}-\tau\left(\boldsymbol{X}_i;t^*\right)\right\}^2, \nonumber
\end{eqnarray}
where $Y_{RL}^*=\frac{I\left(T>t^*\right)-\hat{S}(t^*|\boldsymbol{X})}{A-\hat{e}(\boldsymbol{X})}$ is the pITE and the learner-specific weight is $w^M=\left(A-\hat{e}(\boldsymbol{X})\right)^2$.

Table \ref{tab:metalearner} summarizes the construction of pITE $Y^*$ and method-specific weight $w^M$ for the six pseudo-outcome-based meta-learners we introduced.

\begin{table}[ht]
\caption{pITE $Y^*$ and weight $w^M$ construction for six meta-learners}
\label{tab:metalearner}
    \centering
    \renewcommand{\arraystretch}{2}
    \scalebox{0.86}{
    \begin{tabular}{c|c|c}
    \hline
         Meta-learner & Weight ($w^M$) & Pseudo ITE ($Y^*$)  \\
    \hline
    \multirow{2}{10em}{X-learner}&   & $Y^*_{1,XL}=I(T>t^*)-\hat{S}_0(t^*|\boldsymbol{X})$\\ & 1 & $Y^*_{0,XL}=\hat{S}_1(t^*|\boldsymbol{X})-I(T>t^*)$ \\
    \hline
    \multirow{1}{10em}{M-learner} & 1 & $Y^*_{ML}=\frac{AI(T>t^*)}{\hat{e}(\boldsymbol{X})}-\frac{(1-A)I(T>t^*)}{1-\hat{e}(\boldsymbol{X})}$\\
    \hline
    \multirow{2}{10em}{DR-learner}&  & $Y^*_{DR}=\frac{A-\hat{e}(\boldsymbol{X})}{\hat{e}(\boldsymbol{X})\left(1-\hat{e}(\boldsymbol{X})\right)}\left(I(T>t^*)-\hat{S}_A(t^*|\boldsymbol{X})\right)$ \\
    & 1 & $+\hat{S}_1(t^*|\boldsymbol{X})-\hat{S}_0(t^*|\boldsymbol{X})$ \\
    \hline
     \multirow{1}{10em}{D-learning} & $(2A-1)\frac{A-\hat{e}(\boldsymbol{X})}{4\hat{e}(\boldsymbol{X})\left(1-\hat{e}(\boldsymbol{X})\right)}$ & $Y^*_{DL}=2(2A-1)I(T>t^*)$ \\
    \hline
     \multirow{1}{10em}{DEA-learning} & $(2A-1)\frac{A-\hat{e}(\boldsymbol{X})}{4\hat{e}(\boldsymbol{X})\left(1-\hat{e}(\boldsymbol{X})\right)}$ & $Y^*_{DEA}=2(2A-1)\left(I(T>t^*)-\hat{S}(t^*|\boldsymbol{X})\right)$ \\
    \hline
     \multirow{1}{10em}{R-learner} & $\left(A-\hat{e}(\boldsymbol{X})\right)^2$ & $Y^*_{RL}=\frac{I(T>t^*)-\hat{S}(t^*|\boldsymbol{X})}{A-\hat{e}(\boldsymbol{X})}$\\
    \hline\multicolumn{3}{l}{\small The subscription in $Y^*$ denotes the abbreviation for each meta-learner.}\\ 
    \end{tabular} }
    \renewcommand{\arraystretch}{2}
\end{table} 

Among the six meta-learners, the pITE in X-learner has an actual meaning of ITE. The propensity score does not balance the covariate distributions between two treatment arms, rather, it is used to re-weight CATE estimates on each treatment arm if the sample sizes are unbalanced. M- and DR-learner balance the covariate distribution through re-weighting outcomes. 
D-learner re-weights covariates rather than outcomes, which does not require estimating the conditional survival functions. DEA-learner uses the residual rather than the outcome itself to augment the efficiency. R-learner eliminates the spurious effect by controlling the correlations between $e(\boldsymbol{X})$ and $S(t^*|\boldsymbol{X})$, which is also called Neyman orthogonality. It also has a doubly robust feature. The choice of meta-learners primarily depends on the study characteristics and the complication of the heterogeneity in the covariate space between two treatment arms. These properties will be discussed in more detail in Section \ref{sec:simu}.

\subsubsection{Algorithm to estimate CATE}
\label{sec:method_metalearner_algorithm}
For the proposed pseudo-outcome-based meta-learners, we describe the two-step algorithm to estimate CATE in
Algorithm \ref{algorithm:framework}.
\begin{algorithm}
\caption{\textbf{Pseudo-outcome meta-learner to estimate CATE in survival data}}
\label{algorithm:framework}

\textbf{Input:} Covariate $\boldsymbol{X}$, treatment $A$, pre-specified time $t^*$, observed outcome $I(T>t^*)$.

\textbf{Step1. pITE construction}

(1) Estimate nuisance parameters: propensity score $e(\boldsymbol{X})$ and outcome regression models $S_A(t^*|\boldsymbol{X})$ or $S(t^*|\boldsymbol{X})$.

Use samples who are not censored by $t^*$, denoted as the complete data with sample size $n^o$ to do the following: 

(2) Construct pITE $Y^*$ using a meta-learner approach described in Table \ref{tab:metalearner}.

(3) Construct the learner-specific weight $w^M$ in Table \ref{tab:metalearner}. 

(4) Estimate the inverse probability censoring weight $w^C=\frac{1}{P(C>\min (T, t^*)|\boldsymbol{X},A)}$.

\textbf{Step 2. Pseudo-outcome regression}

Minimize the squared error loss on complete data: $\frac{1}{n^o} \sum_{i=1}^{n^o} w_i^C w_i^M\left(Y_i^*-\tau(\boldsymbol{X}_i;t^*)\right)^2$ with $Y^*$ and weights $w^M$ and $w^C$ constructed from Step 1.

\textbf{Output:} $\hat{\tau}(\boldsymbol{x};t^*)$.

\end{algorithm}

In Step 1, parametric, semiparametric, or nonparametric methods can be used to estimate nuisance parameters. For example, random survival forests (RSF) \cite{RSF}, BAFT, or deep neural network survival model (DNNSurv) \cite{DNNSurv} can be considered to estimate $S_A(t^*|\boldsymbol{X})$ or $S(t^*|\boldsymbol{X})$. The propensity score $e(\boldsymbol{X})$ can be estimated using logistic regressions or random forests (RF) \cite{RF}. The censoring mechanism can be estimated by the KM method or the Cox model. 
In Step 2, either model-based regressions or machine learning methods such as Lasso \cite{lasso1996_Tibshirani} or RF can be applied for pseudo-outcome regressions.

\subsection{Identification of predictive variables contributing to CATE}
\label{sec:method_discover_variable}

In addition to CATE predictions, clinicians are also interested in knowing the variables that impact the predictions. This motivates us to examine the ability of the pseudo-outcome-based meta-learners to identify predictive variables for CATE. 

We will use SHapley Additive exPlanation (SHAP) to quantify the importance of the variable. Lundberg and Lee \cite{SHAP_2017} proposed the SHAP value to explain the contribution of each covariate for an individual prediction. 
Specifically when estimating CATE, for each individual with covariate $\boldsymbol{x}_i=\{x_{ij}\}_{j=1}^p$, the SHAP value of the $j$th covariate, denoted as $v_{ij}$, measures the contribution of $x_{ij}$ to each individual prediction $\hat{\tau}(\boldsymbol{x}_i;t^*)$ by quantifying the effect of removing the rest of the covariates in all coalitions. To save computational load, we used the KernelSHAP version, implemented through the objective function proposed in Martínez \cite{interpretCATE_Javier2021Cambridge}.

\subsection{Subgroup discovery}
\label{sec:method_subgroup}
After estimating CATE and identifying predictive variables, the next step is to delineate the treatment heterogeneity and identify subgroups with enhanced treatment effects (noted as ``beneficial subgroups''). In most existing research, individuals are classified into a beneficial subgroup if their $\hat{\tau}(\boldsymbol{X_i};t^*)$ is greater than a pre-specified value (e.g., 0), which does not demonstrate the distribution of the treatment heterogeneity across the study population. Here, we propose a data-adaptive strategy that dynamically examines the distribution of estimated CATE in the entire study population to help select beneficial subgroups.

Zhao \cite{MTD_Zhao_JASA2013} proposed to calculate the mean treatment difference (MTD) to evaluate the performance of the fitted model for CATE estimation. In this paper, we borrow their approach for the survival outcome. Specifically, let $H(\hat{\tau}(\boldsymbol{x};t^*))$ represent the empirical cumulative distribution function of $\hat{\tau}(\boldsymbol{x};t^*)$. Then, $\hat{\tau}_q=H_q^{-1}(\hat{\tau}(\boldsymbol{x};t^*))$ represents the $q$-th percentile of $\hat{\tau}(\boldsymbol{x};t^*)$, i.e.,  $q=H(\hat{\tau}_q)$. We calculate the MTD as the KM difference between treatment $A=1$ and control $A=0$ at time $t^*$ for the patients who have predicted CATE greater than $\hat{\tau}_{q}$, denoted as $\text{MTD}_{1-q}(t^*)$ with:
\begin{eqnarray}
\label{eq:MTD}
    \text{MTD}_{1-q}(t^*)=\prod_{t\leq t^*}\left\{1-\frac{\sum_{i=1}^n d_{i,q}^{(1)}(t)}{\sum_{i=1}^n R_{i,q}^{(1)}(t)}\right\} -\prod_{t\leq t^*}\left\{1-\frac{\sum_{i=1}^n d_{i,q}^{(0)}(t)}{\sum_{i=1}^n R_{i,q}^{(0)}(t)}\right\},
\end{eqnarray}
where $d_{i,q}^{(a)}(t)=\sum_{i=1}^n I(T_i=t, \delta_i=1, A_i=a, \hat{\tau}(\boldsymbol{x}_i;t^*) \geq \hat{\tau}_q)$ and $R_{i,q}^{(a)}(t)=\sum_{i=1}^n I(T_i \geq t, A_i=a, \hat{\tau}(\boldsymbol{x}_i;t^*) \geq \hat{\tau}_q)$. For example, when choosing 10\% as the spacing interval for the percentiles $q=90\%, 80\%,\dots,10\%$, $\text{MTD}_{1-90\%}(t^*)$ is the KM difference at $t^*$ between two treatment groups in patients who have predicted CATE greater than $\hat{\tau}_{90\%}$, denoted as the ``top 10\% subgroup".
Then we plot $\text{MTD}_{1-q}(t^*)$ versus $1-q$ and compare this MTD curve with the overall MTD value (defined as the KM difference between the treatment and control groups at $t^*$ for all patients). Well-predicted CATEs lead to a decreasing MTD curve towards the overall MTD. The beneficial subgroups are the percentiles with MTD values substantially greater than the overall MTD. Given that higher predicted CATE corresponds to larger treatment benefits, this data-adaptive strategy will help select the top percentage of patients with the most substantial treatment benefits, characterized by the largest predicted CATEs. We will further illustrate it in the real data analysis (Figure \ref{fig:KM_top_quant_R_MTD_split7_1st10splits} in Section \ref{sec:realdat_application}).

\section{Simulation Studies}
\label{sec:simu}
In this section, we performed comprehensive simulations under observational study settings to assess the performance of six meta-learners in estimating CATE and the ability to identify predictive variables for CATE. Results under RCT design will be discussed in Section \ref{sec:simu_results_summary}.

\subsection{Simulation designs}
\label{sec:simu_design}

We generated two observational studies under two types of treatment allocation: a balanced design with equal sample sizes in each treatment arm and an unbalanced design with more samples in the control arm $A=0$. Considering 10 baseline covariates, $X_1, X_2, X_3, X_4$ and $X_5$ were generated independently from $N(0,1)$ and $X_6, X_7, X_8, X_9$ and $ X_{10}$ were binary covariates created by the sign of $\tilde{X}_j \sim N(0,1)$, where $j=6,7, \dots, 10$. Under the balanced design, treatment allocation was determined by the propensity score: $logit(e(\boldsymbol{X}))=0.15-0.8X_1+0.5X_2-0.9X_3-0.9X_4+0.6X_6+0.7X_7-0.8X_8-0.9X_9$. Under the unbalanced design, 30\% of patients were assigned to the treatment arm $A=1$ using the propensity score: $logit(e(\boldsymbol{X}))=1.2(-1.2-0.8X_1+0.5X_2-0.9X_3-0.9X_4+0.6X_6+0.7X_7-0.8X_8-0.9X_9)$. In both settings, we observed a moderate overlapping distribution of the propensity scores between the two treatment groups (Web Figure 1). 
Potential survival times $T(1)$ and $T(0)$ were simulated from a Weibull regression with at least 50\% subjects with a positive CATE: $T(A|\boldsymbol{X})=\lambda_A\{ \frac{-log(U)}{exp\{f_A(\boldsymbol{X})\}}\}^\frac{1}{\eta}$, where $U \sim Unif[0,1]$, the shape parameter $\eta=2$ and the scale parameters $\lambda_0=18$ and $\lambda_1=20$. Here, $f_A(\boldsymbol{X})=b(\boldsymbol{X})+h(\boldsymbol{X})*A$ introduces the source of HTE. $b(\boldsymbol{X})$ is a baseline function shared between two treatment arms, which indirectly contributes to CATE; $h(\boldsymbol{X})$ introduces the treatment-specific effect, which directly contributes to CATE. 
Censoring time was independently simulated from an exponential distribution with 30\% censoring rate. We describe three specific scenarios in the box below. In S1 and S2, we considered a linear function $h(\boldsymbol{X})$ with three covariates directly contributing to CATE, while in S3, we considered a non-linear $h(\boldsymbol{X})$. For the base function $b(\boldsymbol{X})$, we considered a linear case in S1 and a non-linear case in S2 and S3.
\begin{algorithm}
\caption*{\textbf{simulation scenarios}}
\label{eq:scenarios}
     \textbf{S1:
     } 
     $b(\boldsymbol{X})=0.25X_1+0.7X_3+0.5X_6+0.4X_7+0.3X_{10}$, \nonumber  

     \qquad $h(\boldsymbol{X})=-0.1X_2+0.95X_5-0.6X_8$. \nonumber

     \textbf{S2: 
     }
     $b(\boldsymbol{X})=0.35e^{X_1}+0.4X_2^2+0.7\sin(X_3)-0.2X_5 +0.6\sin(X_6)+0.5X_7+0.45X_1X_8-0.15X_{10}$, \nonumber 

     \qquad $h(\boldsymbol{X})=-0.1X_2+0.95X_5-0.6X_8$. \nonumber 

     \textbf{S3: 
     }
     $b(\boldsymbol{X})=0.35e^{X_1}+0.4X_2^2+0.7\sin(X_3)-0.2X_5 +0.6\sin(X_6)+0.5X_7+0.45X_1X_8-0.15X_{10}$, \nonumber 

     \qquad $h(\boldsymbol{X})=-0.2X_1^2-0.25X_2X_3+0.2e^{X_5}-0.3X_7-0.4X_3X_8$. \nonumber
\end{algorithm}

In our previous study \cite{metalearnerSurv_Bo2023}, we conducted a comparison of the X-learner and T-learner under RCT settings using three distinct machine learning methods: RSF, BAFT, and DNNSurv. The analysis revealed that BAFT exhibited more biases and larger RMSEs in CATE estimates, in contrast to RSF and DNNSurv, which demonstrated smaller biases and RMSEs. RSF showed advantage in computational speed as compared to DNNSurv. 
Therefore, in this simulation study, instead of comparing various machine learning methods for estimating nuisance parameters for all meta-learners, we only chose RSF to estimate $S_A(t^*|\boldsymbol{X})$ or $S(t^*|\boldsymbol{X})$ and RF to estimate $e(\boldsymbol{X})$ in Step 1 of Algorithm \ref{algorithm:framework}. 
The KM estimator was used to estimate the censoring probability in $w^C$. 
For Step 2 of Algorithm \ref{algorithm:framework}, we implemented RF for the pseudo-outcome regression. We did not use sample splitting for nuisance parameter estimation and objective function optimization in Algorithm \ref{algorithm:framework}, as Curth et al. \cite{curth2021nonparametric} showed no substantial improvement in the empirical performance when splitting the samples. 
As a comparison, we also estimated the CATE using a non-meta-learner method, i.e., the CSF method. 
True Weibull regression models were also fitted for each treatment arm to construct CATE estimates using the difference of two estimated survival probabilities ($\hat{S}_1(t^*|\boldsymbol{X})-\hat{S}_0(t^*|\boldsymbol{X})$). 
We did 100 runs for each scenario with $n=1,000$ samples in training data and $N=10,000$ samples in test data. $t^*$ was chosen at the median survival time. 
Sensitivity analysis were conducted at different $t^*$ time points. 

\subsection{Evaluation metrics}
\label{sec:simu_eval} \hfill \break
\indent \textbf{Prediction performance.} To examine the ability of each method to uncover treatment heterogeneity, we evaluate their performance within subgroups. We use subgroup-based root mean squared error (RMSE), namely, binned-RMSE, to evaluate predictions \cite{metalearner_surv_LiangyuanHu2021}. We divide predicted CATE values into $Q$ bins based on the ordering of true CATE with $Q=50$. Bias and binned-RMSE are defined as: 
\begin{eqnarray}
     \text{Bias} =\frac{1}{N}\sum\limits_{i=1}^{N}(\hat{\tau}(\boldsymbol{x}_i;t^*)-\tau(\boldsymbol{x}_i;t^*)), \quad \nonumber
     \text{RMSE} = \frac{1}{Q}\sum\limits_{q=1}^{Q}\sqrt{\frac{1}{N_q}\sum\limits_{i=1}^{N_q}(\hat{\tau}(\boldsymbol{x}_i;t^*)-\tau(\boldsymbol{x}_i;t^*))^2},\nonumber
\end{eqnarray}
where $\tau(\boldsymbol{x}_i;t^*) $ is the true CATE and $\hat{\tau}(\boldsymbol{x}_i;t^*)$ is the predicted CATE for sample $i$, and $N_q$ is the sample size of the test data in the $q$-th bin.

We also evaluate the performance of each method in predicting patients with beneficial treatment effects (i.e., a positive CATE $\tau(\boldsymbol{x};t^*)>0$). We use overall accuracy (ACC), positive predictive value (PPV), negative predictive value (NPV), sensitivity, specificity, and F-score as metrics. The true positive (TP) (or false positive (FP)) is defined as the number of patients with true CATE $>$ 0 (or $\leq$ 0) and predicted CATE $>$ 0. The true negative (TN) and false negative (FN) are defined similarly. We further define: $\text{ACC}=\frac{\text{TP}+\text{TN}}{\text{N}}$, $\text{PPV}=\frac{\text{TP}}{\text{TP}+\text{FP}}$, $\text{NPV}=\frac{\text{TN}}{\text{TN}+\text{FN}}$, $\text{sensitivity}=\frac{\text{TP}}{\text{TP}+\text{FN}}$, $\text{specificity}=\frac{\text{TN}}{\text{TN}+\text{FP}}$, and $\text{F-score}=\frac{2}{\text{PPV}^{-1}+\text{sensitivity}^{-1}}$.

\textbf{Evaluation of variable importance.} To examine the ability of each method in identifying predictive variables directly contributing to CATE (i.e., variables in $h(\boldsymbol{X})$), we use the ``attribution score" \cite{benchmark_HTE_interpret_crabbé2022}, which is a summery-level metric calculated on the test data, define as
\begin{eqnarray}
\label{eq:simu_eval_attr}
    \text{Attr}_{\text{pred }}=\frac{1}{\left|{D}_{\text {test }}\right|} \sum_{i \in {D}_{\text {test }}} \frac{\sum_{j \in {V}_h}\left|v_{ij}\right|}{\sum_{j=1}^p\left|v_{ij}\right|}, \nonumber
\end{eqnarray}
where ${D}_{\text {test }}$ denotes the test data, $\left|{D}_{\text {test }}\right|$ denotes the sample size of test data, ${V}_h$ denotes variables in function $h(\boldsymbol{X})$, and $v_{ij}$ is the KernelSHAP value for the $j$-th covariate in $i-$th individual defined in Section \ref{sec:method_discover_variable}.

\subsection{Simulation results}
\label{sec:simu_results}

\subsubsection{Results under the balanced design of observational studies}
\label{sec:simu_results_predict}
\begin{figure}[ht]
\centering
\includegraphics[width=5in]{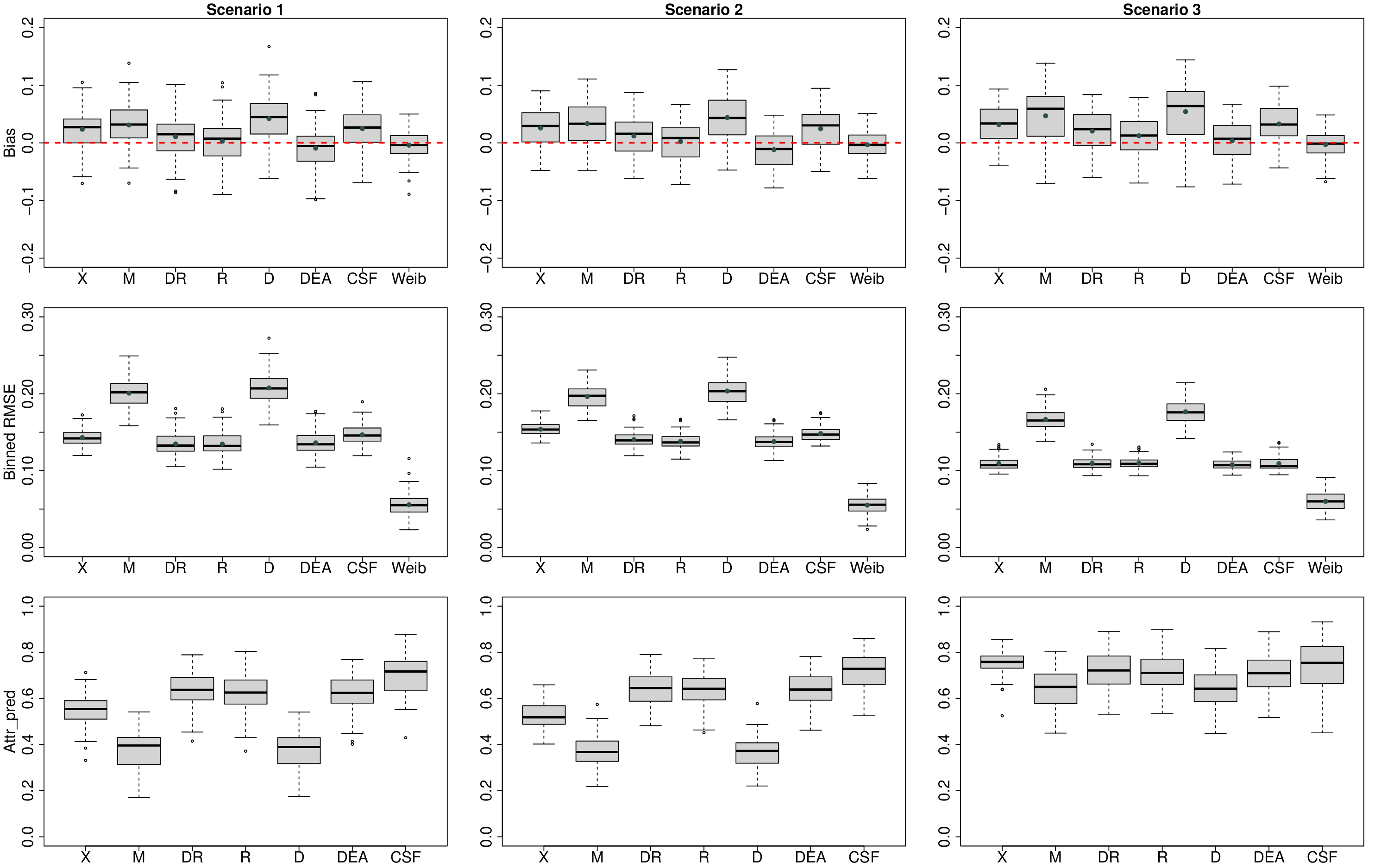}
\caption{The top two panels are the prediction performance under the balanced design of observational studies at $t^*=$ median survival time. The bottom panel is the attribution score of variables directly contributing to CATE. X, M, DR, R, D, and DEA are six meta-learners; CSF: causal survival forests; Weib: true Weibull model. To save computational load, we calculated the attribution score of each method on 100 test samples.}
\label{fig:simu_obs_balanced_weak_lowdim_pseudorf}
\end{figure}

R- and DEA- learners generally perform the best under the balanced design of the observational studies, showing minimal biases and small RMSEs across all three scenarios (the upper and middle panels of Figure \ref{fig:simu_obs_balanced_weak_lowdim_pseudorf}). However, X-, M-, D- learners and CSF show obviously large biases in all three scenarios. DR-learner shows comparable RMSEs to R- and DEA- learners but shows more biases, especially when the source of HTE gets complex (scenario 3). 
Notably, CSF and X-learner tend to predict more positive CATEs than the truth, especially when the source of HTE gets complex (e.g., CSF tends to predict all CATE positive in around 40\% simulations under scenario 3). The bottom panel of Figure \ref{fig:simu_obs_balanced_weak_lowdim_pseudorf} shows the attribution scores of each method, which measures the importance of variables directly contributing to HTE (e.g., $X$'s from $h(\boldsymbol{X})$ in Section \ref{sec:simu_design}). R-, DEA- and DR-learner show the highest attribution scores in all three scenarios. Despite X-learner and CSF showing relatively high attribution scores in all scenarios, they produce large biases in predictions. 

Web Table 1 presents the accuracy measures of each method in three scenarios. R- and DEA- learners show the best performance in overall accuracy, PPV, specificity and F-score. 
DR-learner also shows comparable accuracy compared to R- and DEA- learners in these measures. X-learner and CSF show larger NPVs and sensitivities, but they tend to produce large biases.  
In summary, under the balanced design of observational studies, the R- and DEA-learner perform the best in all aspects.

\subsubsection{Results under other designs}
\label{sec:simu_results_summary}
In general, we observe similar performance results in terms of biases, RMSEs, and prediction accuracy under the unbalanced observational studies and the RCT studies (when treatment was randomly assigned with ($P(A=1|\boldsymbol{X})=0.5$) at median survival time. R- and DEA- learners perform the best except that the DEA-learner shows slightly larger biases in scenarios 1 and 2 (Web Figure 2) compared to the R-learner under the unbalanced design of observational studies. Under the RCT design, all six meta-learners and CSF show minimal biases, with DR-, R- and DEA- learners showing the smallest RMSEs (Web Figure 3). CSF and X-learners still tend to predict more positive CATEs than the truth. In examining the ability of each method in identifying predictive variables directly contributing to CATE, DR-, R-, and DEA- learners still show the highest attribution scores across all scenarios and designs (Web Figure 4).

We further chose the 75\% percentile of survival times (as $t^*$) to conduct a sensitivity analysis. Our results show that the six meta-learners and CSF are all robust to the choice of times in terms of prediction accuracy, as we observe similar results compared to $t^*=$median survival time in all scenarios and designs (Web Figure 5 to 7).

To conclude, for balanced observational studies, we recommend R- and DEA-learner, while for unbalanced observational studies, we recommend R-learner. For RCTs, we recommend DR-, R-, and DEA-learner.

\section{Application to Pediatric Asthma Care under COVID-19 Disruption}
\label{sec:realdat_application}
In this section, we analyze the HTE of WAAP on time-to-ED return with COVID-19 disruption. The analysis determined HTE for each patient and identified vulnerable subgroups, offering insights for pediatric asthma surveillance under a disruptive public health event.

\subsection{Estimation of CATE of WAAP on time-to-ED return}
\label{sec:realdat_application_predict}
In this section, we estimated the CATE of WAAP on time-to-ED return, where CATE is defined as the difference in ED return-free probabilities between treated (WAAP) and control (no WAAP) groups up to $t^*=$ one year given covariates. Given the unbalanced allocation of WAAP (25\% with WAAP vs. 75\% without WAAP at the index ED visit), we applied R-learner for estimating CATE. The covariates include age, sex, race, the existence of chronic diseases, the existence of acute respiratory diseases, influenza vaccination, hospitalization, and pandemic period. 
CATEs were estimated through a 5-fold cross-validation where four folds of the data were used for training (both steps in Algorithm \ref{algorithm:framework}) and the remaining one fold were used for predictions. We applied RSF to estimate the conditional survival probability and RF to estimate the propensity score of getting WAAP. 
We then calculated the MTD metric (\ref{eq:MTD}) for patients in each top $(1-q)$ subgroup. 

\begin{figure}[ht]
\centering
\includegraphics[width=4.5in]{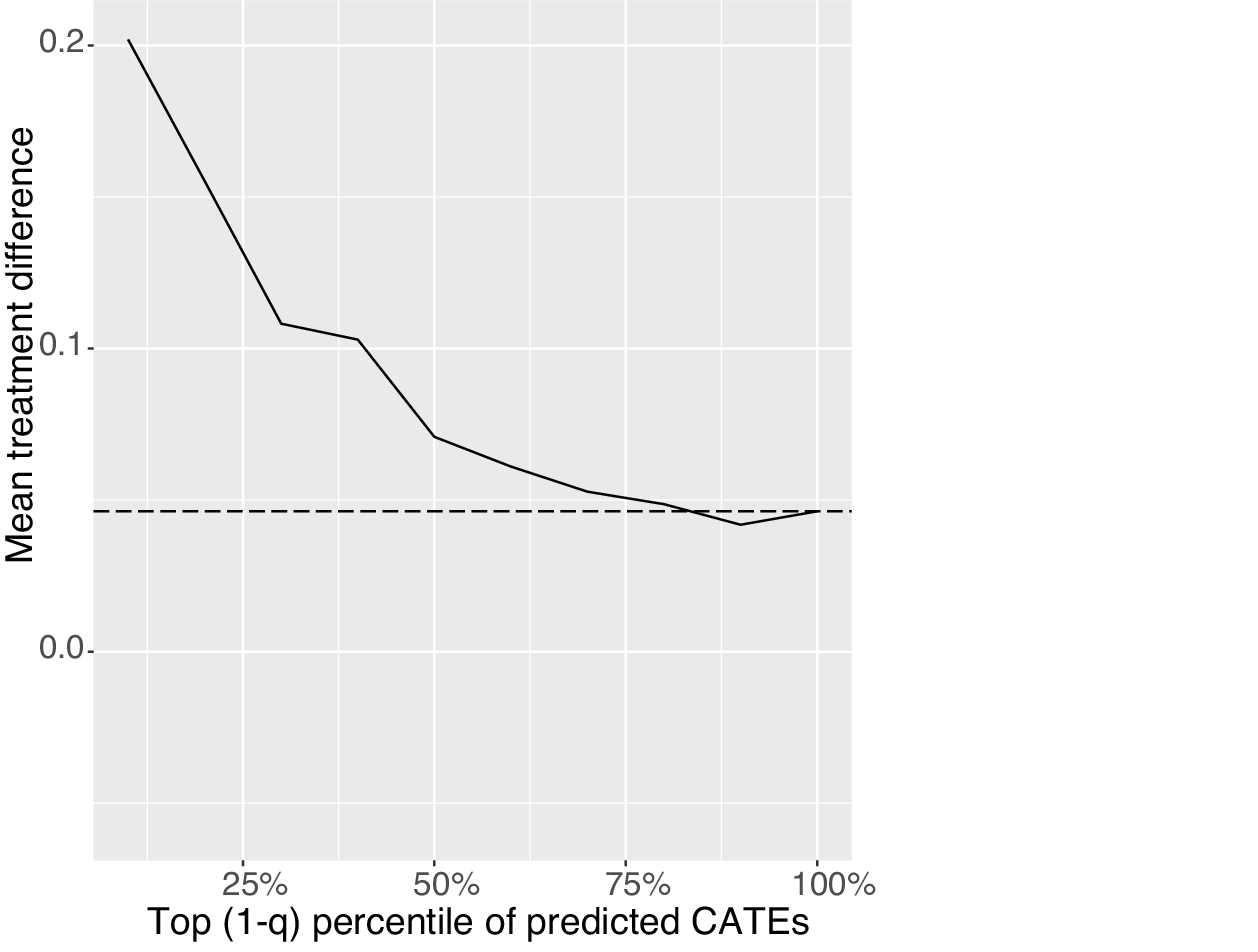}
\caption{Mean treatment difference (MTD) over top $1-q$ subgroup of patients for $q=90\%, 80\%, \dots, 10\%$, where $q$'s are the percentiles of predicted CATEs using R-learner. The horizontal dashed line is the overall MTD in all patients.}
\label{fig:KM_top_quant_R_MTD_split7_1st10splits}
\end{figure}

Figure \ref{fig:KM_top_quant_R_MTD_split7_1st10splits} shows the MTD curve over the top $(1-q)$ subgroup of patients, where $q=90\%, 80\%, \dots, 10\%$ denotes the $q$-th percentile of predicted CATEs. The dashed line indicates a small positive average effect of WAAP (overall MTD=0.046). The MTD curve shows a positive benefit from WAAP for all patients. Notably, the MTD curve shows a decreasing trend with $1-q$ increases and reaches the overall MTD value with top 75\% of patients. Specifically, the top 10\% have the largest benefit from WAAP with an MTD of 0.21, and the top 20\% also have an MTD of 0.15, which is considerably bigger than the overall MTD. This MTD curve clearly demonstrates heterogeneous WAAP effects exist in this study population.

\subsection{Identification of predictive variables}
\label{sec:realdat_application_vip}
Next, we identified predictive variables that contribute to CATEs by computing VIPs using the KernelSHAP metric as described in Section \ref{sec:method_discover_variable}. Similar to the real data application in paper that proposed SHAP value \cite{SHAP_2017}, we identified the top predictors by looking at the center of the distribution of absolute KernelSHAP values for each covariate (Web Figure 8). The top predictors that we identified include the influenza vaccination, pandemic period, and patient age. Subsequently, race, sex, and hospitalization are secondary predictors of CATEs, with race exhibiting comparatively greater predictability than the other two variables.

\subsection{Subgroup analysis}
\label{sec:realdat_application_subgroup}
\begin{figure}[ht]
\centering
\includegraphics[width=4.5in]{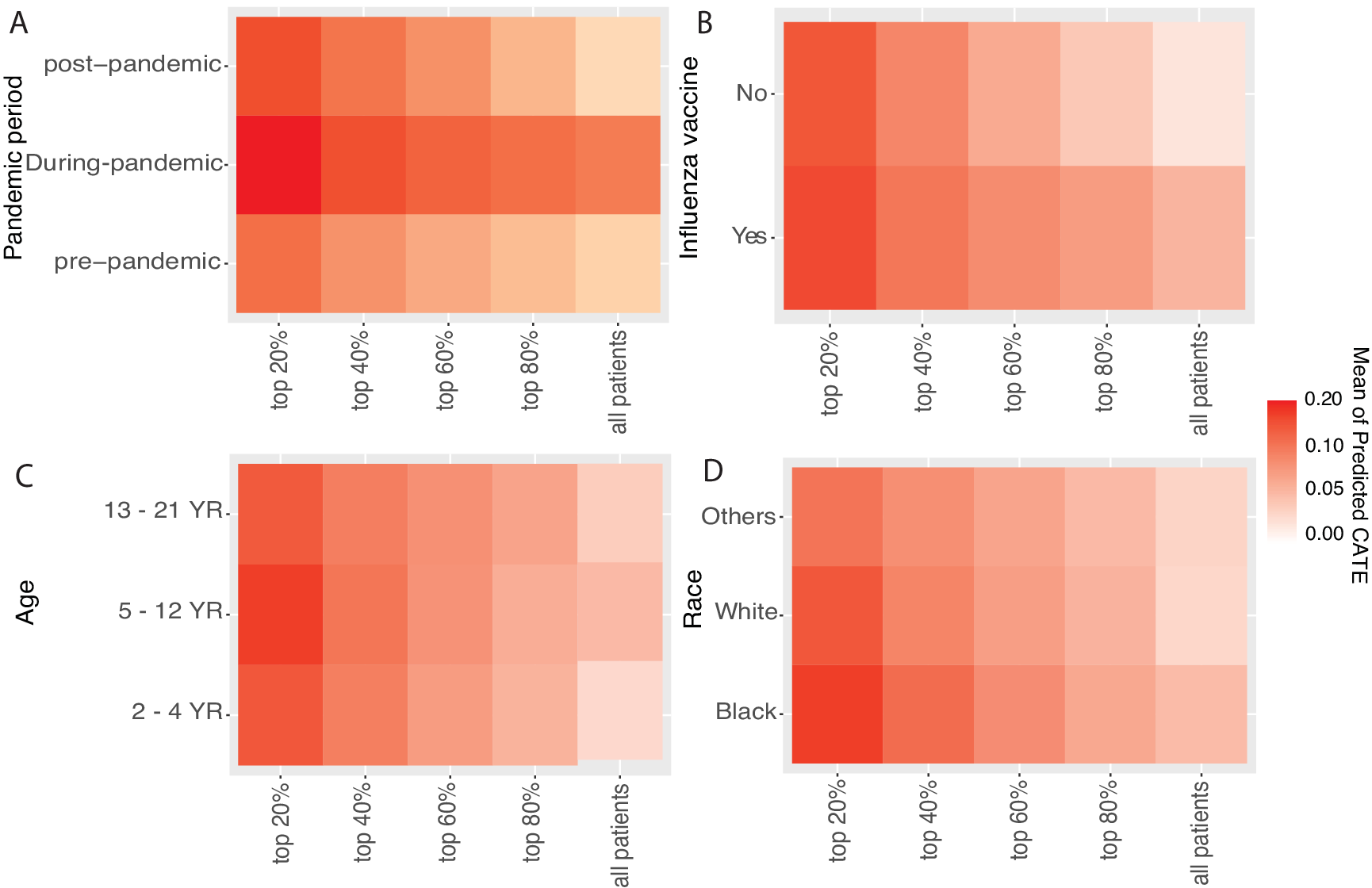}
\caption{Mean of the predicted CATE values in each top $(1-q)$ subgroup by top predictors. The color represents the averaged CATE under each top $(1-q)$ subgroup. }
\label{fig:asthma_subgroup_aveCATE_by_cov_split7_1st10splits}
\end{figure}

In Figure \ref{fig:asthma_subgroup_aveCATE_by_cov_split7_1st10splits}, we further examined the differential subgroup effects by visualizing the mean of the predicted CATE for each top 
$(1-q)$ subgroup with spacing interval being 20\%, grouped by the top four predictors identified in Section \ref{sec:realdat_application_vip}. 
A darker color represents a larger mean CATE. Notably, patients whose index ED visits were during the pandemic consistently show larger CATE values compared to the pre- and post-pandemic periods in all percentiles (Figure \ref{fig:asthma_subgroup_aveCATE_by_cov_split7_1st10splits}.A), suggesting enhanced effects of WAAP during the pandemic in terms of delaying the ED return. Patients who had influenza vaccination consistently exhibit larger CATE values across all percentiles compared to patients who did not have influenza vaccination (Figure \ref{fig:asthma_subgroup_aveCATE_by_cov_split7_1st10splits}.B). A noteworthy heterogeneous effect of WAAP is observed across age groups, where elementary school-age patients (5-12 years) consistently show larger CATE than other patients, especially in the top 20\% and 40\% percentiles (Figure \ref{fig:asthma_subgroup_aveCATE_by_cov_split7_1st10splits}.C). Finally, black patients gain enhanced protection from WAAP compared to white and other race patients (Figure \ref{fig:asthma_subgroup_aveCATE_by_cov_split7_1st10splits}.D), suggesting the importance of protecting black patients by offering WAAP, especially when a disruptive event happens.

Our analysis shows that WAAP demonstrates enhanced protection against pediatric asthma exacerbation in delaying ED returns, especially during and after the pandemic. The findings demonstrate the importance of prioritizing those vulnerable subgroups (e.g., black patients and elementary school-age patients) in terms of mitigation protocols when future pandemics or disruptive events happen by ensuring that they have access to WAAP. Moreover, advocating influenza vaccination is also important to mitigate asthma exacerbation. Implementing these recommendations will help develop a precision surveillance strategy to help mitigate adverse impacts on asthma care from future disruptive events.

\section{Conclusion and Discussion}
\label{sec:discuss}
In this paper, we proposed a pseudo-outcome-based framework for analyzing HTE in survival data, which includes a list of pseudo-outcome-based meta-learners to estimate CATE, the kernelSHAP metric to identify predictive variables, and a data-adaptive approach to select beneficial subgroups. We implemented six meta-learners and evaluated their performance in terms of CATE prediction and the ability to identify contributing variables to CATE in various observational studies and RCT settings. Among the six meta-learners, we recommend DEA- and R-learner, which produce minimum biases and small RMSEs in CATE estimation and also effectively identify variables directly contributing to CATE in survival data. 

We applied R-learner to the pediatric asthma EHR data and estimated the HTE of WAAP on delaying time-to-ED return. We identified vulnerable subgroups with enhanced benefits from WAAP and characterized the key predictors contributing to the enhanced treatment effects. Our analysis suggests prioritizing black and young patients in terms of mitigation protocols when future public health disruptive events happen to ensure they have access to WAAP. These results provide valuable insights for developing a precision surveillance strategy for pediatric asthma care.

One limitation of this work is the beneficial subgroups are discovered through a post-hoc analysis. To enhance the interpretability of the meta-learners, one may further consider integrating interpretable algorithms to estimate CATE and identify subgroups simultaneously, such as the interpretable algorithm proposed by Wan et al.\cite{Rulefit_KeWan2023_SIM,wan2023survival_arxiv}. Another limitation is that we assumed non-informative censoring; although is valid for observational studies in many cases, it is worthwhile to develop meta-learners when informative censoring exists. While our aim of this work focuses on the estimation of CATE and subgroup discovery rather than the inference of CATE, the latter remains a crucial and challenging topic in this domain. One potential direction is to develop new methods using conformal inference to construct prediction intervals, as demonstrated in traditional survival analysis (not causal) \cite{ConformalSurvival_Candes2023} and causal ITE predictions in continuous outcomes \cite{ConformalITE_Lei2021}. 
We are currently investigating these two directions -- interpretable subgroups and conformal inference for survival CATE.

\section*{Supporting information}
Web tables and figures referenced in
Section~\ref{sec:data_explore}, \ref{sec:simu} and \ref{sec:realdat_application} can be found in the online version of the article at the publisher’s website. The proposed methods are implemented in R. The key functions can be found in Github \\
\url{https://github.com/nab1779321/metalearner_shap_surv}.

\bibliographystyle{plain}
\bibliography{bibliography}

\begin{thebibliography}{10}

\bibitem{cross_section_survey_AAP_Alkhthlan2021}
Abdullah Alkhthlan.
\newblock The effects of an asthma action plan and asthma self-efficacy on asthma control.
\newblock Master's thesis, Georgia State University, 2021.

\bibitem{metalearnerSurv_Bo2023}
Na~Bo, Yue Wei, Lang Zeng, Chaeryon Kang, and Ying Ding.
\newblock A meta-learner framework to estimate individualized treatment effects for survival outcomes.
\newblock {\em Journal of Data Science}, pages 1--19, 2024.

\bibitem{AsthmaEducation_RCT_BROWN2006}
Michael~D. Brown, Mathew~J. Reeves, Karen Meyerson, and Steven~J. Korzeniewski.
\newblock Randomized trial of a comprehensive asthma education program after an emergency department visit.
\newblock {\em Annals of Allergy, Asthma \& Immunology}, 97(1):44--51, 2006.

\bibitem{ConformalSurvival_Candes2023}
Emmanuel Candès, Lihua Lei, and Zhimei Ren.
\newblock {Conformalized survival analysis}.
\newblock {\em Journal of the Royal Statistical Society Series B: Statistical Methodology}, 85(1):24--45, 01 2023.

\bibitem{DEAlearning2017}
Shuai Chen, Lu~Tian, Tianxi Cai, and Menggang Yu.
\newblock A general statistical framework for subgroup identification and comparative treatment scoring.
\newblock {\em Biometrics}, 73(4):1199--1209, 2017.

\bibitem{benchmark_HTE_interpret_crabbé2022}
Jonathan Crabbé, Alicia Curth, Ioana Bica, and Mihaela van~der Schaar.
\newblock Benchmarking heterogeneous treatment effect models through the lens of interpretability, 2022.

\bibitem{CSF_JRSSb2023}
Yifan Cui, Michael~R Kosorok, Erik Sverdrup, Stefan Wager, and Ruoqing Zhu.
\newblock {Estimating heterogeneous treatment effects with right-censored data via causal survival forests}.
\newblock {\em Journal of the Royal Statistical Society Series B: Statistical Methodology}, 02 2023.
\newblock qkac001.

\bibitem{survITE_NeurIPS2021_Curth}
Alicia Curth, Changhee Lee, and Mihaela van~der Schaar.
\newblock Survite: Learning heterogeneous treatment effects from time-to-event data.
\newblock In M.~Ranzato, A.~Beygelzimer, Y.~Dauphin, P.S. Liang, and J.~Wortman Vaughan, editors, {\em Advances in Neural Information Processing Systems}, volume~34, pages 26740--26753. Curran Associates, Inc., 2021.

\bibitem{curth2021nonparametric}
Alicia Curth and Mihaela van~der Schaar.
\newblock Nonparametric estimation of heterogeneous treatment effects: From theory to learning algorithms, 2021.

\bibitem{BAFT_Henderson2018_biostatistics}
Nicholas~C Henderson, Thomas~A Louis, Gary~L Rosner, and Ravi Varadhan.
\newblock {Individualized treatment effects with censored data via fully nonparametric Bayesian accelerated failure time models}.
\newblock {\em Biostatistics}, 21(1):50--68, 07 2018.

\bibitem{IPW1952_Horvitz_Thompson}
D.~G. Horvitz and D.~J. Thompson.
\newblock A generalization of sampling without replacement from a finite universe.
\newblock {\em Journal of the American Statistical Association}, 47(260):663--685, 1952.

\bibitem{clusteredCATEsurv_Hu2022}
Liangyuan Hu, Jiayi Ji, Ronald~D. Ennis, and Joseph~W. Hogan.
\newblock A flexible approach for causal inference with multiple treatments and clustered survival outcomes.
\newblock {\em Statistics in Medicine}, 41(25):4982--4999, 2022.

\bibitem{HTE_mediansurv_ML_Hu2021}
Liangyuan Hu, Jiayi Ji, and Fan Li.
\newblock Estimating heterogeneous survival treatment effect in observational data using machine learning.
\newblock {\em Statistics in Medicine}, 40(21):4691--4713, 2021.

\bibitem{metalearner_surv_LiangyuanHu2021}
Liangyuan Hu, Jiayi Ji, and Fan Li.
\newblock Estimating heterogeneous survival treatment effect in observational data using machine learning.
\newblock {\em Statistics in Medicine}, 40(21):4691--4713, 2021.

\bibitem{RSF}
H.~Ishwaran and U.B. Kogalur.
\newblock Random survival forests for r.
\newblock {\em R News}, 7(2):25--31, October 2007.

\bibitem{kennedyDR2022}
Edward~H. Kennedy.
\newblock {Towards optimal doubly robust estimation of heterogeneous causal effects}.
\newblock {\em Electronic Journal of Statistics}, 17(2):3008 -- 3049, 2023.

\bibitem{pnas_metalearner2019}
Soren~R. Kunzel, Jasjeet~S. Sekhon, Peter~J. Bickel, and Bin Yu.
\newblock Metalearners for estimating heterogeneous treatment effects using machine learning.
\newblock {\em dings of the National Academy of Sciences of the United States of America}, 116(10):4156--4165, 2018.

\bibitem{ConformalITE_Lei2021}
Lihua Lei and Emmanuel~J. Candès.
\newblock {Conformal Inference of Counterfactuals and Individual Treatment Effects}.
\newblock {\em Journal of the Royal Statistical Society Series B: Statistical Methodology}, 83(5):911--938, 10 2021.

\bibitem{RF}
Andy Liaw and Matthew Wiener.
\newblock Classification and regression by randomforest.
\newblock {\em R News}, 2(3):18--22, 2002.

\bibitem{SHAP_2017}
Scott~M. Lundberg and Su-In Lee.
\newblock A unified approach to interpreting model predictions.
\newblock In {\em Proceedings of the 31st International Conference on Neural Information Processing Systems}, NIPS'17, page 4768–4777, Red Hook, NY, USA, 2017. Curran Associates Inc.

\bibitem{interpretCATE_Javier2021Cambridge}
Javier~Abad Martínez.
\newblock Interpretability for conditional average treatment effect estimation.
\newblock Master's thesis, University of Cambridge, 8 2021.

\bibitem{Rlearner2021}
X~Nie and S~Wager.
\newblock {Quasi-oracle estimation of heterogeneous treatment effects}.
\newblock {\em Biometrika}, 108(2):299--319, 09 2020.

\bibitem{rubin1974}
Donald~B Rubin.
\newblock Estimating causal effects of treatments in randomized and nonrandomized studies.
\newblock {\em Journal of Educational Psychology}, 66(4):688--701, 1974.

\bibitem{Var_ChildrenAsthmaED_EHR_Shechter2019}
Jesse Shechter, Angkana Roy, Sara Naureckas, Christopher Estabrook, and Nivedita Mohanty.
\newblock Variables associated with emergency department utilization by pediatric patients with asthma in a federally qualified health center.
\newblock {\em Journal of community health}, 44(5):948--953, 10 2019.
\newblock Copyright - Journal of Community Health is a copyright of Springer, (2019). All Rights Reserved; Last updated - 2023-03-01.

\bibitem{Neyman1990}
J.~Splawa-Neyman, D.M. Dabrowska, and T.~Speed.
\newblock On the application of probability theory to agricultural experiments. essay on principles.
\newblock {\em Statistical Science}, 5(4):465--472, 1990.

\bibitem{DNNSurv}
Tao Sun, Yue Wei, Wei Chen, and Ying Ding.
\newblock Genome-wide association study-based deep learning for survival prediction.
\newblock {\em Statistics in Medicine}, 39:4605--4620, 2020.

\bibitem{ED_WAAP_RCT_HighMorbidity_children_Teach2006}
Stephen~J. Teach, Ellen~F. Crain, Deborah~M. Quint, Michelle~L. Hylan, and Jill~G. Joseph.
\newblock {Improved Asthma Outcomes in a High-Morbidity Pediatric Population: Results of an Emergency Department–Based Randomized Clinical Trial}.
\newblock {\em Archives of Pediatrics \& Adolescent Medicine}, 160(5):535--541, 05 2006.

\bibitem{Dlearning_TianLu2014}
Lu~Tian, Ash~A. Alizadeh, Andrew~J. Gentles, and Robert Tibshirani.
\newblock A simple method for estimating interactions between a treatment and a large number of covariates.
\newblock {\em Journal of the American Statistical Association}, 109(508):1517--1532, 2014.
\newblock PMID: 25729117.

\bibitem{lasso1996_Tibshirani}
Robert Tibshirani.
\newblock Regression shrinkage and selection via the lasso.
\newblock {\em Journal of the Royal Statistical Society: Series B (Methodological)}, 58(1):267--288, 1996.

\bibitem{Rulefit_KeWan2023_SIM}
Ke~Wan, Kensuke Tanioka, and Toshio Shimokawa.
\newblock Rule ensemble method with adaptive group lasso for heterogeneous treatment effect estimation.
\newblock {\em Statistics in Medicine}, 42(19):3413--3442, 2023.

\bibitem{wan2023survival_arxiv}
Ke~Wan, Kensuke Tanioka, and Toshio Shimokawa.
\newblock Survival causal rule ensemble method considering the main effect for estimating heterogeneous treatment effects, 2023.

\bibitem{metalearner_survival_casestudy_Xu2023}
Yizhe Xu, Katelyn Bechler, Alison Callahan, and Nigam Shah.
\newblock Principled estimation and evaluation of treatment effect heterogeneity: A case study application to dabigatran for patients with atrial fibrillation.
\newblock {\em Journal of Biomedical Informatics}, 143:104420, 2023.

\bibitem{CATEsurvTutorialBookChap_Xu2022_inbook}
Yizhe Xu, Nikolaos Ignatiadis, Erik Sverdrup, Scott Fleming, Stefan Wager, and Nigam Shah.
\newblock {\em Handbook of Matching and Weighting Adjustments for Causal Inference}, chapter Chapter 21.Treatment Heterogeneity with Survival Outcomes.
\newblock Chapman and Hall/CRC, 2023.

\bibitem{MTD_Zhao_JASA2013}
Lihui Zhao, Lu~Tian, Tianxi Cai, Brian Claggett, and L.~J. Wei.
\newblock Effectively selecting a target population for a future comparative study.
\newblock {\em Journal of the American Statistical Association}, 108(502):527--539, 2013.
\newblock PMID: 24058223.

\bibitem{TMLE_survival_Zhu_JBI2020}
Jie Zhu and Blanca Gallego.
\newblock Targeted estimation of heterogeneous treatment effect in observational survival analysis.
\newblock {\em Journal of Biomedical Informatics}, 107:103474, 2020.

\end{thebibliography}

\label{lastpage}

\end{document}